\pageno=0
\nopagenumbers
\footline= { \ifnum\pageno>0 \hss\tenrm\folio\hss \fi}
\magnification=\magstep1 

\def \e{{\rm e}}
\def \i{{\rm i}}
\def \m{\textstyle {1\over 2}}
\def \tz{\textstyle {1\over 3}}
\null
\vskip 2pc
{\centerline {SELF-SIMILARITY OF THE NEGATIVE BINOMIAL}}
{\centerline {MULTIPLICITY DISTRIBUTIONS}}
\vskip 2pc

\centerline{G. Calucci and D. Treleani}
\vskip 2pc
\centerline{\it Dipartimento di Fisica Teorica dell'Universit\`a and INFN.}
\centerline{\it Trieste, I 34014 Italy}
\vskip 2pc

{\centerline {ABSTRACT}}
\vskip 2pc
{\midinsert
\narrower
The negative binomial distribution is self similar: If the spectrum over the
whole rapidity range gives rise to a negative binomial, in absence of 
correlation and if the source is unique, also a partial range in rapidity
gives rise to the same distribution. The property is not seen in experimental
data, which are rather consistent with the presence of a number of independent
sources. When multiplicities are very large self similarity might be used to
isolate individual sources is a complex production process.

\endinsert}
\vfill \eject
One of the first basic evidences observed in the field of many-particle 
production and nuclear collisions is the distribution of the multiplicity 
of the produced particles. Multiplicity distributions are measured both by
looking at the whole spectrum of the produced particles and by looking only
at a restricted segment, typically a rapidity interval. Both for theoretical
and experimental reasons, one of the favorite parametrization of the 
multiplicity distribution [1], also in different rapidity intervals [2], is the
negative binomial distributions ({\it NB}). 
A very detailed discussion of the experimental evidences, of the 
interpretations and also of the formalism
used to deal with this kind of problems has been recently published [3]. 
In the case of a generic distribution the relation between the multiplicities
of a restricted part of the spectrum and those arising from the whole spectrum
is not trivial. In the present note we point out that for the {\it NB}, on the
contrary, a peculiar self-similarity property holds between the distributions
obtained from different intervals of the spectrum.
\par
We find convenient to make use of the generating 
functional formalism to deal with this kind of problems [4,5,6].
Let $ W_n(\xi_1,\dots,\xi_n)$ be the normalized multiparticle exclusive 
distributions:
$$ \sum_n \int W_n(\xi_1,\dots,\xi_n)d\xi_1,\dots,d\xi_n =1 .\eqno (1)$$
The variables $\xi$
can have different meanings and also represent more than one physical
parameters. In high-energy
collisions $\xi$ could represent the rapidity $y$ and the transverse momentum,
if the distributions refer to incoming partons $\xi$ could represent the
fractional longitudinal momentum $x$ and the impact parameter.
The distributions may be obtained in the usual way from a generating functional
 $\cal Z $:
$$ W_n(\xi_1,\dots,\xi_n)={1\over{n!}}{\delta \over{\delta J(\xi_1)}} \dots 
{\delta \over{\delta J(\xi_n)}} {\cal Z}[J]\,|_{J=0} \eqno (2)$$
and the normalization is expressed by $ {\cal Z}[1]=1$.
Sometimes it will be useful also to use an unrenormalized generator 
${\cal G}$ with  ${\cal Z}[J]={\cal G}[J] /{\cal G}[1]$
\par
The probability of producing $n$ particles, in any configuration, is 
evidently given by:
 $$ p_n= \int W_n(\xi_1,\dots,\xi_n)d\xi_1,\dots,d\xi_n
={1\over{n!}}\Bigl[\int{\delta \over{\delta J(\xi)}}d\xi\Bigr]^n 
{\cal Z}[J]\,|_{J=0}=$$
$$={1\over{n!}}\Bigl[{\partial \over {\partial \lambda }}\Bigr]^n 
{\cal Z}[J+\lambda 1]\,|_{J=0,\lambda=0}={1\over{n!}}
\Bigl[{\partial \over {\partial \lambda }}
\Bigr]^n {\cal Z}[\lambda 1]\,|_{\lambda=0}={1\over{n!}}
\Bigl[{\partial \over {\partial \lambda }}\Bigr]^n
z(\lambda )\,|_{\lambda=0}. \eqno (3)$$
\par
Let us now consider the situation where the interval in which the variables 
$\xi$ lie is divided into two parts. Then for a particular choice of these 
variables it results:
$$ W_n(\xi_1,\dots,\xi_n)= W_r(\xi'_1,\dots,\xi'_r)W_s(\xi''_1,\dots,\xi''_s)$$
with $r+s=n$.
Taking into account all the possible choices of $\xi'$ and $\xi''$ it results:
$$ W_r(\xi'_1,\dots,\xi'_r) W_s(\xi''_1,\dots,\xi''_s)=
{1\over{r!}}{\delta \over{\delta J(\xi'_1)}} \dots {\delta \over{\delta 
J(\xi'_r)}}
{1\over{s!}}{\delta \over{\delta J(\xi''_1)}} \dots 
{\delta \over{\delta J(\xi''_s)}} {\cal Z}[J]\,|_{J=0}.$$
If we sum over all configurations in $\xi''$ the distributions in $\xi'$ are:
$$ W_r(\xi'_1,\dots,\xi'_r)\cdot 
\sum \int W_s(\xi''_1,\dots,\xi''_s)d\xi''_1,\dots,d\xi''_s$$
A set of semi-inclusive distributions are obtained in this way since everything
 referring to the
variables $\xi''$ is not observed. The generator of these new distributions is
 ${\cal Z}'={\cal Z}[J'+\Theta '']$
where $J'$ has as argument only $\xi'\;i.e.\;J'(\xi'')=0$, $\Theta ''$ is 1 for 
$\xi=\xi''$ and 0 for $\xi=\xi'$, $\Theta '$ is 1 for $\xi=\xi'$ and 0 for 
$\xi=\xi''$. The probability of finding $n$ particles in the observed part of 
the spectrum is then:
$$p'_n={1\over{n!}}
\Bigl[{\partial \over {\partial \lambda }}\Bigr]^n 
{\cal Z}[\lambda \Theta '+\Theta '']\,|_{\lambda=0}={1\over{n!}}
\Bigl[{\partial \over {\partial \lambda }}\Bigr]^n
z'(\lambda )\,|_{\lambda=0}. \eqno (4)$$
\vskip 1pc
Two particular cases of interest are:
\par
 The Poissonian distribution, which is obtained by defining:
$$\;{\cal U}=\int J(\xi)\cdot D(\xi) d\xi \quad ,\quad
\tilde u=\int D(\xi) d\xi \quad ,\quad
{\cal G}=\e^{{\cal U}[J]}\quad ,\quad {\cal G}_1=\e^{\tilde u}$$
and finally: ${\cal Z}=\e^{{\cal U}[J]-\tilde u}$
\par
If one looks only at the spectrum in $\xi'$ by integrating over $\xi''$, the
new generator is
$${\cal G'}=\e^{{\cal U}[J(\xi')]+{\cal U}{[\Theta '']}}\,,$$
since $\tilde u={\cal U}[\Theta ']+{\cal U}[\Theta'']$
it results ${\cal Z'}={\cal Z}$.
\par
The {\it NB} distribution, whose generating functional is:
$$f({\cal U})=
{[1-{\cal U}]^{-k} \over {[1-\tilde u]^{-k}}} \eqno (5) $$
while the generator of the semi-inclusive spectra in $\xi'$ is
$${{\{1-{\cal U}[J(\xi')]-{\cal U}{[\Theta '']\}^{-k}} 
\over {[1-\tilde u]^{-k}}}}.$$
\par 
This corresponds to a pure redefinition of ${\cal U}$ since one gets the new 
generator by going from 
  ${\cal Z}=f\{{\cal U}\}$ to
  ${\cal Z'}=f\{{\cal U}/(1-{\cal U}[\Theta ''])\}$. This means that the 
NB is transformed into a NB, with the same 
exponent as the original one. Clearly, in both cases, the mean multiplicity is
changed.
\par
The generating function of the multiplicity distribution in this case is  
explicitly given as
$$ z(\lambda)={{[1-\lambda u'-u'']^{-k}} 
\over {[1-\tilde u]^{-k}}}, \eqno (6)$$
or, after defining $r=u'/(1-\tilde u)$, in a different and sometimes more
convenient form        
$$ z(\lambda)=[1+(1-\lambda)r]^{-k}. \eqno (6')$$
In term of these parameters one gets for the mean multiplicity $\bar n=kr$ and
for the dispersion $D^2=kr(r+1)$.
\par
A survey of other kinds of one-body distributions shows that this property of
self-similarity if only a part of the spectrum is 
observed is quite unlikely,\footnote *{$e.g.\,$the {\it NB} is a
particular case of a hypergeometric distribution, but
a generic hypergeometric distribution does not have this kind of 
self-similarity } one may therefore wonder whether this property is 
peculiar of the {\it NB} distribution, with the Poissonian distribution as a 
limiting case,or it is also found in other cases.
\par 
It will be shown that in the simplest conditions the property of self-similarity
is unique of the {\it NB} distribution. In this case one can give for the non 
normalized generating functional the representation
 ${\cal G}=g({\cal U});$ the probabilities $p'$, eq (4),
can be obtained from a generating function $g(\lambda u'+u'')$
where $$u'=\int D(\xi')d\xi'\quad ,\quad u''=\int D(\xi'')d\xi'' \quad ,\quad  
u'+u''=\tilde u. \eqno (7)$$
The invariance of the functional form of the distribution, when considering 
only limited parts of the spectrum is expressed as:
$g(x+y)=N(y) g(x\cdot f(y))$ because in this way the relation $p'_n=c^np_n/C$
is produced, and this property can be expressed by saying that the distribution
remains the same.
The arbitrary normalization $g(0)=1$, which is always possible, gives
$N(y)=g(y)$. So finally: 
 $$g(x+y)=g(y) g(x\cdot f(y)) \eqno (8)$$
By taking the first and the second derivative with respect to $x$ and setting
then $x=0$, two differential equations for $g(y)$ are obtained
$$\dot g(y)=\dot g(0) g(y)f(y)\quad \ddot g(y)=\ddot g(0) g(y) f(y)^2. 
 \eqno (8')$$
It follows then
$$ g(y) \ddot g(y)=R \dot g(y)^2 \qquad{\rm with}:
\qquad R=\ddot g(0) /\dot g(0)^2\;.$$
With the usual position $$g(y)=\exp\Bigl[\int_0^yq(w)dw\Bigr]\;,$$
which ensures the correct normalization $g(0)=1$, the equation becomes
 $$\dot q(y)=(R-1)q(y)^2 \eqno (8'')$$
The solution of eq. (8'') is:
$$q(y)=[(1-R)y+S]^{-1}\;.$$
Redefining the constants as $k=1/(R-1)$ and $u=(R-1)/S$ one obtains
 $$g_u(y)=[1-uy]^{-k}\;, \eqno (9)$$
This expression is the generating functional of a binomial distribution
whose exponent is, in general, not integer. The meaning of the function $g(\xi)$
requires that it be positive together with all its derivatives in the origin,
this certainly happens if the exponent is negative, $i.e.\; R>1$ and the 
parameter $u$ is positive. A different possibility is given by positive integer
exponent and negative $u$.
This corresponds, however to a distribution with only a finite number of
terms.
\par
The two differential equations eq.(8') are not completely equivalent to the 
functional relation eq.(8), but they follow from it. The conclusion is that 
the self similarity implies the {\it NB} (which could be not sufficient) but it
has already shown that the {\it NB} implies the self similarity, so the
two properties are equivalent.
The generating functional of the {\it NB} is more conveniently expressed
by writing $g_u(\lambda)$ as $g_1(\lambda u)$ and suppressing from 
now on the index $1$; the normalized distribution is given by
$z(\lambda)=g(\lambda u)/g(u).$
\par
 The limit $R \to 1$ gives rise to the solution $g(y)=\exp[y/S]$ $i.e.$
it yields the generating function for a Poissonian distribution.
\vskip 1pc
The experimental evidences and their elaboration [7,8] show that the {\it NB} 
distribution holds well for different intervals of observed rapidity but that
the parameters present strong variations. Real world does not shows the sharp
self-similarity property discussed above. The actual analysis was done in a
frame where ${\cal Z}=f\{{\cal U}\}$ so that case genuine two-body 
correlation were absent.

When correlations are present the relation between exclusive and 
semi-inclusive distribution is more complicated and there is no obvious reason
for the self similarity to hold.
However this way does not seems too promising: either the effect of the
correlations is so strong that the {\it NB} distribution is destroyed or the 
overall effect is not very important but then the parameters of the {\it NB} 
distribution are changed too little to agree with the experimental evidence. 
An example will be shown in the Appendix.
\par
A more interesting possibility is given by the often considered possibility 
[1,2,8] of considering multiple sources in the rapidity range.
\par
Let us consider a simple case where a source extends in rapidity from $y_o$
to $y_1$ and another source is present from  $y_1$ to $y_2$: when we observe
the produced particles in a rapidity range that ends at $y_f < y_1$ then the
second source in inactive, the parameter $r$ grows with $y_f$ and does the
multiplicity, the parameter $k$ stays evidently constant. When  $y_f$ goes
beyond $y_1$ the first source is frozen ($r$ has attained its final value) 
and the second gives a contribution still growing with $y_f$. 
The generating function is now
$$ z(\lambda)=[1+(1-\lambda)r]^{-k}\cdot [1+(1-\lambda)r_f]^{-k}
 \eqno (10)$$
and does not yield a NB distribution. One could force the function $z(\lambda)$
to become a NB-generating function:
$$ z_e(\lambda)=[1+(1-\lambda)r_e ]^{-k_e} \eqno (11) $$
by defining the equivalent parameters in such a way that multiplicity and
dispersion acquire the correct values. The prescription is expressed through the
auxiliary parameter $\rho = r_f/r$.
$$r_e =r (1+\rho ^2)/(1+\rho)\quad ,\quad k_e =
k (1+\rho )^2 /(1+\rho ^2).$$
In order to explore how good this representation it is useful to calculate the
higher central momenta $\mu_s =<(n-<n>)^s>$. The third central momentum 
indicates that the worst situation is produced for $r_f \approx \tz r $ and a 
similar indication is obtained by examining the fourth cumulant [9]
$\;\kappa_4 =\mu_4 -3D^2$; in this situation the error cannot exceed 12\%.
One can also examine in details the individual 
distribution of the multiplicity produced respectively by the generating
functions eq.(10) and eq.(11); it results that the approximation is better
than it could seems at first sight because large deviations between the two
series of numbers is found for multiplicities very large, typically a
discrepancy of the order 12\% arises for multiplicities of the order of 25 
which gives sizeable contributions to the higher momenta but are not very 
relevant in the analysis of the data; for values form 6 to 9, where the
maximum of the production rate lies the difference is less than 1\%. These
values are obtained for $r_f \approx \tz r $, in other cases the 
discrepancy is definitely smaller.
 Anyhow, without dwelling furthermore on a particular form of
approximation the conclusion that we try to draw is that a number of
sources each of them giving rise to a strict {\it NB} distribution within a
definite range of rapidity yields a distribution not very different when
taken over the whole rapidity range.
\par
If one would try to construct a model for high-energy particle production which
implies sources extended in rapidity, one would like to determine the extension
in rapidity of the individual sources.
 A qualitative examination of the distributions associated to events 
with 2,3,4 jets suggests that the extension of the individual source cannot be
the same in the different families of events but, better, that it
is larger in the 2-jets events an becomes narrower and narrower in passing to
the configurations with 3 and 4 jets.
The extension in $y$ of the sources cannot become too narrow, if this should
happen so that the number of sources grows too much, the generating function
would approach the corresponding expression for the Poissonian distribution. 
\par
When many sources are active the present description of the multiple production 
acquires many similarities with the "clan" description [8].
On the other hand a feature of the two-source model discussed previously is that
 it is
possible that only a part of the source is active, the description
we start with is in fact differential in $y$. The model lacks informations
on the transverse dynamics which certainly enters also in the multiplicity
distributions. In fact the total multiplicity is larger when the jets
number is larger [7,8], in the description here presented this would require
that more than one source is active in the same rapidity interval, what looks
very artificial if we neglect the transverse degrees of freedom but becomes 
quite natural when transverse degrees of freedom are taken into account
The model of multiple sources just described is anyhow still rather rough, in
particular one would not expect sharp beginning and a sharp end for the 
rapidity range where the source is active. The present accuracy of the 
experimental data, however, does not allow to discriminate the actual model from
different possibilities. A further point is that the sources have been taken
as equivalent: the presence of internal quantum numbers, which may affect the
production mechanism [10] have not been taken into account.
\par
A rather general feature, associated with the presence of different sources 
ordered in rapidity is a weak, long-range correlation in rapidity among the 
particles.
 This may be seen in the following way: the generating functional eq.(5) is 
substituted by a product
$$f({\cal U})=
\prod_n {[1-{\cal U}_n]^{-k} \over {[1-\tilde u_n]^{-k}}}; \eqno (5') $$
every factor $n$ acts in a different range of rapidity.
If the two particles lie in the same rapidity interval, two body distribution 
is 
$$ D(\xi_1 ,\xi_2)=A k(k+1) D(\xi_1) D(\xi_2)[1-\tilde u_{n'}]^2\;;$$
whereas
if the two particles lie in different rapidity intervals it results
$$ D(\xi_1 ,\xi_2)=A k^2 D(\xi_1) D(\xi_2)[1-\tilde u_{n'}]
[1-\tilde u_{n''}]\;.$$
In both cases $A=\prod_n {[1-\tilde u_n]^k}$.
\par
In conclusion the main points of the present analysis are summarized: The 
success of the {\it NB} in describing the multiparticle distributions supports
the possibility that the {\it NB} is the actual distribution arising from a 
single source. The characterizing property of the {\it NB} is the self 
similarity: if the source is unique, when considering a part of the spectrum
one obtains the same {\it NB} distribution which describes the total spectrum.
The large variation of the {\it NB} parameters as a function of the rapidity
interval in multiparticle production in therefore a strong indication for the 
presence of many sources. The alternative possibility is the presence of 
correlation within a single source. If the distribution in the whole spectrum
is a {\it NB}, correlations most probably produce different distributions
when looking at different parts of the spectrum. On the contrary, as in the
model discussed above, the superposition of different sources, each giving 
rise to a {\it NB} distribution, can easily produce distributions which are
close to a {\it NB} with altered parameters.
\par
Hence one could consider, in high energy processes with very high multiplicity,
to use the self similarity property in order to isolate different sources
which are active in a complex production process. Events could be organized
by considering different topologies $e.g.$ number of jets, impact parameter
(in heavy ion collisions) $etc.$ and one could look at multiplicity 
distributions in different regions of phase space. The individual sources are
isolated when, subdividing further the phase space regions, the corresponding
multiplicity distributions are self similar.
\vskip 2pc
{\bf Acknowledgments}
\vskip 1pc \noindent
This work was partially supported by the Italian Ministry of University and of
Scientific and Technological Research by means of the Fondi per la Ricerca
scientifica - Universit\`a di Trieste.
\vfill
\eject 
{\bf {Appendix}}
\vskip 1pc
In this appendix only the two-body correlations are studied, so
beyond the linear term ${\cal U}[J]=\int J(\xi) D(\xi) d\xi$ also a term
${\cal V}[J,J]=\m \int C(\xi_1,\xi_2)J(\xi_1)J(\xi_2) d\xi_1 d\xi_2$ is used
with the condition ${\cal V}[1,1]=0$. Then a generating functional, with these
restrictions, can be expressed as:
${\cal Z}=g({\cal U}[J],{\cal V}[J,J])/g({\cal U}[1])$ so that the corresponding
generating function for the multiplicities is 
$z(\lambda)=g(\lambda \tilde u)/g(\tilde u)$.
If one looks only to one part of the spectrum, the one can define the 
corresponding multiplicities according to eq (4) and the result is:
 $$z(\lambda )=g(\lambda u'+u'',\lambda^2 v'+2\lambda \bar v+v'')/g(\tilde u).
\eqno(A1)$$
The terms $\tilde u,u',u''$ have been already defined in eq.(5), the definition 
of 
the $v$-terms, where the symmetry of $C$ has been used, is:                   
               $$v'=\m \int C(\xi'_1,\xi'_2) d\xi'_1 d\xi'_2 \quad
            \bar v=\m \int C(\xi'_1,\xi''_2) d\xi'_1 d\xi''_2 \quad
               v''=\m \int C(\xi''_1,\xi''_2) d\xi''_1 d\xi''_2 ;\eqno (A2)$$
the initial condition ${\cal V}[1,1]=0$ is translated into $v'+2\bar v+v''=0$
which will be used in order to eliminate the term $\bar v$.
 \par
Now one can look to particular cases and the most interesting seems to be
precisely a distribution which produces a {\it NB} multiplicity when 
integrated over the whole spectrum but contains two-body correlations. 
The simplest form in which the generating functional may be written is:
$$ {\cal Z}=f({\cal U})=
{[1-{\cal U}-{\cal V}]^{-k} \over {[1-\tilde u]^{-k}}}\eqno (A3) $$
and when only a part of the spectrum is observed and the rest is integrated 
over the generating function of the multiplicity is:
$$z(\lambda)={ {[1-\tilde u]^{k}}\over{[1-(u''+\lambda u')-(\lambda^2 v'+
2\lambda \bar v +v'')]^{k}}}\;\eqno (A4).$$
It is useful to write the same expression in a more compact form $i.e.$:

$$z(\lambda)=N\cdot [1-\lambda a-\lambda^2 b]^{-k},\eqno (A5)$$
having defined 
 $$
    a={{u'-v'-v''}\over {1-u''-v''}} \quad ,\quad
    b={{v'}\over {1-u''-v''}} \quad ,\quad
      N=[1-a-b]^{k}=\Bigl[{{1-\tilde u}\over {1-u''-v''}}\Bigr]^{k} \eqno (A6)$$
The new expression for the multiplicity distribution is now obtained by 
expanding $z(\lambda )$, as given in eq (A5), in powers of $\lambda$; the
result is
$$z(\lambda)=N\cdot \sum_n (\i \lambda \sqrt {b})^n
             C^{(k )}_n (\i a/2 \sqrt {b}), \eqno (A7)$$
where  $C^{(k )}_n$ represents the Gegenbauer polynomial [9] of index
${k }$ and order $n$.
This kind of expansion does not look very transparent, anyhow from the explicit
form of the Gegenbauer polynomials it is easily seen that every term of the
sum is real, as obviously it must be; it is also straightforward to verify
that when the effect of the correlations vanishes, so $v', v'', b$ go to zero,
the usual binomial distribution is recovered.
\par
If the correlations are present but not very strong the terms $v$
will be small and one can perform an expansion in $b$.
To the first order in $b$ the expression of $z(\lambda)$ is

$$ \eqalign {
z(\lambda)=N\cdot &\bigl[[1-\lambda a]^{-k} 
(1-2k b/a^2)+\cr 
&\bigl([1-\lambda a]^{-k +1}+[1-\lambda a]^{-k -1}\bigr)
k b/a^2 \bigr].} \eqno (A8)$$

With this expansion the original binomial distribution is reproduced
, with some small correction for the parameter, but other satellite
binomial distributions arise, whose exponent is shifted by $\pm 1\,$, so
that the distance from the original distribution increases with the power 
of the of small parameter representing the effect of the correlations.
\par 
Also in presence of correlation there is the limiting relation between the
NB and the Poissonian distribution. In formal way this may be
obtained through the definitions:
$$ {\cal U}={\cal P}/k ,\quad {\cal V}={\cal Q}/k ,\quad
   u=p/k ,\quad  v=q/k $$
then in the limit $k \to \infty$ out of eq.s (A3,A4) it results
$$ {\cal Z}=\exp[{\cal P}+{\cal Q}-p_1]$$
$$ z(\lambda )=\exp [-(p'+q'')+\lambda (p'-q'-q'') +\lambda^2 q'] $$
and for what concerns eq (A7) one can use the limiting expressions of the 
Gegenbauer polynomials yielding the Hermite polynomials [9].
\par
What appears, beyond the details of the calculations that necessarily refer 
to simplified examples, is that in presence of 
two-body correlations the partial spectra are necessarily different from the 
complete ones: if the correlations play a minor role, then the {\it NB} 
distribution is approximately preserved, but with the too strong result of 
having a constant $k$-parameter, strong correlations may simulate a variable 
$k$-parameter but modify strongly the distribution, which is no longer a
{\it NB}-distribution, not even approximately.
\vfill
\eject

{\bf {References}}
\vskip 1pc
\item {1.}A. Giovannini, L. Van Hove Z. Phys. C30, 213 (1986);
                                    Acta Phys. Polonica B19, 495 and 931 (1988);
                                   Proc XVII  Int.Symposium on Multiparticle
             Dynamics 1987 {\it World Scientific Singapore} p.561

\item {2.} A. Giovannini Proc. int. Conference: Physics
 in collision VI  {\it World Scientific Singapore }1987, p.39

\item {3.}E.A. DeWolf, I.M. Dremin, W. Kittel  Phys. Rep. 270, 1 (1996)
\item {4.}L.S. Brown Phys. Rev. D5, 748 (1972);
         Y. Akiyama, S. Hori Progr. Theor. Phys. 48, 276 (1973)
\item {5.}R. Blankenbecler Phys. Rev. D8, 1611 (1973)
\item {6.}G. Calucci, D. Treleani Int. J. Mod. Phys. A6, 4375 (1991)
\item {7.}P. Abreu {\it et al.} - Delphi Collaboration Z. Phys C 56, 63 (1992)
\item {8.}F. Bianchi, A. Giovannini, S. Lupia, R. Ugoccioni Z. Phys C 58, 71 
 (1993)
\item {9.}M. Abramowitz and I. Stegun - Handbook of mathematical functions 
         (ch. 22) {\it Dover Publications New York} 1965, 1968
\item {10.}C.K. Chew, D. Kiang, H. Zhou Phys. Lett. B186, 411 (1986)
           D. Ghosh {\it et al.} Fizika B6, 37 (1997)
\vfill   
\end
\bye